\documentclass[10pt,letterpaper,preprint,byrevtex,showpacs]{revtex4-1}

\DeclareSymbolFont{lettersA}{U}{pxmia}{m}{it}
\DeclareMathAlphabet{\mathsfsl}{OT1}{cmss}{m}{sl}

\SetSymbolFont{lettersA}{bold}{U}{pxmia}{bx}{it}
\DeclareFontSubstitution{U}{pxmia}{m}{it}
\DeclareSymbolFontAlphabet{\mathfrak}{lettersA}
\DeclareMathSymbol{\piup}{\mathord}{lettersA}{"19}
\DeclareMathSymbol{\iTheta}{\mathalpha}{letters}{2}

\usepackage{amsmath}
\usepackage{amssymb}
\usepackage{graphicx}
\usepackage{epstopdf}
\usepackage{bbm}
\usepackage{mathrsfs}
\usepackage{xcolor}
\usepackage[colorlinks=true,pdfstartview=FitV,linkcolor=blue,citecolor=blue,urlcolor=blue]{hyperref}

\makeatletter

\newcommand{\Rmnum}[1]{\expandafter\@slowromancap\romannumeral #1@}
\makeatother

\newcommand{\blue}[1]{\textcolor{blue}{#1}}

\begin{document}

\title{Giant optical nonlocality near the Dirac point
        in metal-dielectric multilayer metamaterials}

\author{Lei~Sun}
\author{Jie~Gao}
\author{Xiaodong~Yang}
\email[To whom all correspondence should be addressed. \\ Electronic address: ]{yangxia@mst.edu}
\address{Department of Mechanical and Aerospace Engineering,
    Missouri University of Science and Technology,
    Rolla, Missouri 65409, USA}

\begin{abstract}
The giant optical nonlocality near the Dirac point in lossless metal-dielectric
multilayer metamaterials is revealed and investigated through the analysis of the
band structure of the multilayer stack in the three-dimensional $\omega$-$k$ space,
according to the transfer-matrix method with the optical nonlocal effect.
The position of the Dirac point is analytically located in the $\omega$-$k$ space.
It is revealed that the emergence of the Dirac point is due to the degeneracy
of the symmetric and the asymmetric eigenmodes of the coupled surface plasmon
polaritons.
The optical nonlocality induced epsilon-near-zero frequency shift for the
multilayer stack compared to the effective medium is studied.
Furthermore, the giant optical nonlocality around the Dirac point is explored
with the iso-frequency contour analysis, while the beam splitting phenomenon
at the Dirac point due to the optical nonlocal effect is also demonstrated.
\end{abstract}

\maketitle


\section{Introduction}

Dirac point, conical singularity first discovered in the band structure
of graphene \cite{Wallace1947PR},
is the source of many remarkable wave transport properties
in both electronic \cite{Novoselov2005Nat,Zhang2005Nat,Neto2009RMP}
and photonic system \cite{Plihal1991PRB,Sepkhanov2007PRA,Ochiai2009PRB,Diema2010PB,Yannopapas2011PRB},
such as
the long-range flavor current splitting
induced by the giant nonlocality in graphene \cite{Abanin2011Sci},
conical diffraction \cite{Peleg2007PRL},
Bragg-like reflection \cite{Chen2009PRA},
quantum-like Goos-H\"{a}nchen shifts \cite{Chen2009PRA},
and photon \textit{Zitterbewegung} \cite{Zhang2008PRL,Wang2009EPL}
in photonic crystals, because of the linear dispersion relation
in the neighborhood of the Dirac point.
In addition, rigorous analysis based on the tight-binding approximation and group
theory clearly reveals the necessary conditions \cite{Sakoda2012OE1,Sakoda2012OE2}
for the creation of the Dirac point
in the photonic system, including the optical metamaterials, a kind of artificial
composites with periodic subwavelength meta-atoms and tunable electric permittivity
and magnetic permeability.
Particularly, analogous to the effectively massless electron induced by the linear
dispersion near the Dirac point of the electronic system, a Dirac point located
at the center of the Brillouin zone (BZ) in the photonic system and optical metamaterials
implies an effectively zero ``optical mass",
i.e., zero electric permittivity \cite{Wang2009OL,Huang2011NM},
leading significant analogies between the propagation of light in such media
and charge transport in graphene \cite{Bliokh2013ArXiv}.

On the other hand, epsilon-near-zero (ENZ) metamaterials with exotic electromagnetic
properties have been studied in both
theory \cite{Alu2007PRB,Sun2012APL,Sun2013PRB}
and experiment \cite{Roberts2010JN,Alekseyev2010APL,Subramania2012APL,Vesseur2013PRL}
recently.
It is generally thought that the ENZ regime in the domain of the optical
frequency range guarantees the local response described by the effective medium theory (EMT),
since the vanishing effective permittivity implies a long effective
wavelength resulting in a negligible optical size of the period.
However, associated with periodic meta-atom structures in metamaterials,
the variation of the electromagnetic field at the scale of a single
meta-atom will result in the optical nonlocality,
which is studied in both nanolayered \cite{Elser2007APL,Orlov2011PRB,Kruk2012OE,Chebykin2012PRB}
and nanorod \cite{Wurtz2011NN,Pollard2009PRL} metamaterials,
and extended to transformation optics \cite{Castaldi2012PRL}.
Especially, the optical nonlocality is theoretically analyzed
in the metal-dielectric multilayer metamaterials
since the geometrical simplicity allows the exact analytical calculation \cite{Chebykin2012PRB}.
It is worth revealing the mechanism of the optical nonlocality near the Dirac point
in the ENZ regime.

In this work, we demonstrate the giant optical nonlocality near the Dirac point
in lossless metal-dielectric multilayer metamaterials.
Based on the transfer-matrix method for one-dimensional layered photonic crystals,
the band structure of the multilayer stack is fully explored and illustrated in
the three-dimensional $\omega$-$k$ space.
The exact location of the Dirac point in the band structure is located by rigorous
algebraic analysis.
It is revealed that the degeneracy of the symmetric and the asymmetric eigenmodes
of the coupled surface plasmon polaritons (SPPs) induces the emergence of the Dirac
point in the band structure of the multilayer stack, based on the dispersion relation
analysis and the mode analysis around the Dirac point.
Moreover, the locations of the Dirac point and the ENZ response in the $\omega$-$k$ space
for the multilayer stack with the optical nonlocality and the effective medium are also
studied.
The giant optical nonlocality near the Dirac point is revealed by analyzing the
iso-frequency contours (IFCs) in the $k$-space with respect to the EMT
and the transfer-matrix method.
Furthermore, the beam splitting phenomenon induced by the giant nonlocality at the
Dirac point, which is the optical analogue to the long-range flavor current splitting
at the Dirac point in graphene \cite{Abanin2011Sci}, is also demonstrated by numerical simulations.

\section{Band structures and Dirac points of multilayer stack}

Consider a multilayer stack composite of alternating layers of gold ($\mathrm{Au}$)
and alumina ($\mathrm{Al}_{2}\mathrm{O}_{3}$), with the thickness of each layer as
$d_{1}=15\,\mathrm{nm}$ and $d_{2}=35\,\mathrm{nm}$, respectively, as depicted in Fig.~\blue{1(a)}.
The permittivity of $\mathrm{Au}$ is described by the lossless Drude model
$\varepsilon_{1}=\varepsilon_{\infty}-\omega_{p}^{2}/\omega^{2}$, with the
permittivity constant $\varepsilon_{\infty}=5.7$,
and the plasma frequency $\omega_{p}=1.3666\times10^{16}\,\mathrm{rad}/\mathrm{s}$.
Here the material loss of $\mathrm{Au}$ is neglected, since it will suppress the
optical nonlocality of the multilayer stack.
The permittivity of $\mathrm{Al}_{2}\mathrm{O}_{3}$ is $\varepsilon_{2}=2.4$.
The multilayer stack can be regarded as a homogenous effective medium with the
effective permittivity components described by the EMT as
\begin{equation}
\label{eq:emt_x}
    \varepsilon_{x}^{(0)} = \frac{\varepsilon_{1}\varepsilon_{2}(d_{1}+d_{2})}
        {\varepsilon_{1}d_{2}+\varepsilon_{2}d_{1}}
        = \frac{\varepsilon_{1}\varepsilon_{2}}{(1-f_{1})\varepsilon_{1}+f_{1}\varepsilon_{2}},
\end{equation}
\begin{equation}
\label{eq:emt_y}
    \varepsilon_{y}^{(0)} = \varepsilon_{z}^{(0)}
        = \frac{\varepsilon_{1}d_{1}+\varepsilon_{2}d_{2}}{d_{1}+d_{2}}
        = f_{1}\varepsilon_{1}+(1-f_{1})\varepsilon_{2}.
\end{equation}
It is worth mentioning that the EMT, Eqs.~\eqref{eq:emt_x} and \eqref{eq:emt_y},
only take into account the filling ratios of different materials in the multilayer stack,
i.e., the filling ratio of $\mathrm{Au}$ as $f_{1}=d_{1}/(d_{1}+d_{2})$ and the filling
ratio of $\mathrm{Al}_{2}\mathrm{O}_{3}$ as $f_{2}=d_{2}/(d_{1}+d_{2})\equiv 1-f_{1}$,
rather than the actual thickness of each layer.
Hence, EMT is an approximated theory that works well when the period $d=d_{1}+d_{2}$ of
the multilayer stack is much smaller than the wavelength of the propagated electromagnetic
wave within the metamaterial, in the condition of the long wavelength limit.

More precisely, the stack can be regarded as a layered photonic crystal with the
period of $d$ along the $x$-direction, and the period of infinity along the $y$- and
$z$-direction.
Regarding a TM-polarized electromagnetic wave (with non-vanishing components of $E_{x}$, $E_{y}$, and $H_{z}$)
propagating in the $x$-$y$ plane, the band structure of the multilayer stack reads
\begin{equation}
\label{eq:tmm}
\begin{aligned}
    \cos\left(k_{x}(d_{1}+d_{2})\right)
        = & \cos\left( k_{x}^{(1)}d_{1} \right)\cos\left( k_{x}^{(2)}d_{2} \right) \\
          & -\frac{1}{2}
            \left( \frac{\varepsilon_{1}k_{x}^{(2)}}{\varepsilon_{2}k_{x}^{(1)}}
            + \frac{\varepsilon_{2}k_{x}^{(1)}}{\varepsilon_{1}k_{x}^{(2)}} \right)
        \sin\left( k_{x}^{(1)}d_{1} \right)\sin\left( k_{x}^{(2)}d_{2} \right),
\end{aligned}
\end{equation}
according to the transfer-matrix method \cite{Elser2007APL}.
Here $k_{x}^{(1,2)}=\sqrt{k^{2}\varepsilon_{1,2}-k_{y}^{2}}$, where $k=\omega/c$
is the wave vector in free space.
The band structure of the multilayer stack calculated from Eq.~\eqref{eq:tmm} is
displayed in Figs.~\blue{1(b)} and \blue{1(c)} in the first BZ with respect to the period $d$ along
the $x$-direction.
It is clear that the band structure contains three separated bands [Fig.~\blue{1(b)}],
while the band~1 and the band~2 are connected with each other only at two points [Fig.~\blue{1(c)}],
i.e., the Dirac points that will be focused on here.

Regarding a simple dielectric-metal-dielectric, or metal-dielectric-metal multilayer structure,
it is known that the symmetric and the asymmetric SPP modes are the two fundamental propagating
modes in the structure.
Meanwhile, in the band structure of the multilayer stack, the emergence of the two connecting
points, i.e., the Dirac points, implies the mode degeneracy of the symmetric and the asymmetric
eigenmodes.
Therefore, in order to locate the exact position of the two Dirac points in the $\omega$-$k$ space,
the dispersion relation of the SPP along the metal-dielectric interface
$k_{y}/k_{p}=\sqrt{\varepsilon_{1}\varepsilon_{2}/\left(\varepsilon_{1}+\varepsilon_{2}\right)}k/k_{p}$
is substituted into Eq.~\eqref{eq:tmm}, which yields the following equation
\begin{equation}
\label{eq:spp_equ}
    \cos\left(k_{x}(d_{1}+d_{2})\right)
        = \cos\left(k\frac{\varepsilon_{1}d_{1}+\varepsilon_{2}d_{2}}
        {\sqrt{\varepsilon_{1}+\varepsilon_{2}}}\right).
\end{equation}
Here the wave vector $k_{p}=\omega_{p}/c$.
Note that the SPP dispersion relation requires that $\varepsilon_{1}\varepsilon_{2}<0$
and $\varepsilon_{1}+\varepsilon_{2}<0$, leading to the condition of
$\omega/\omega_{p}<1/\sqrt{\varepsilon_{\infty}+\varepsilon_{2}}$,
while a propagating mode in the multilayer stack requires that the wave vector
$k_{x}$ and $k_{y}$ must have real values.
Hence, the only solution of Eq.~\eqref{eq:spp_equ} reads
\begin{equation}
\label{eq:spp_sol}
\left\{
\begin{aligned}
    k_{x}/k_{p} &= 0 \\
    k_{y}/k_{p} &= \sqrt{\frac{\varepsilon_{2}d_{1}d_{2}}
        {(d_{2}-d_{1})(\varepsilon_{\infty}d_{1}+\varepsilon_{2}d_{2})}}
        = \sqrt{\frac{f_{1}(1-f_{1})\varepsilon_{2}}
        {(1-2f_{1})\left(f_{1}\varepsilon_{\infty}+(1-f_{1})\varepsilon_{2}\right)}} \\
    \omega/\omega_{p} &= \sqrt{\frac{d_{1}}{\varepsilon_{\infty}d_{1}+\varepsilon_{2}d_{2}}}
        = \sqrt{\frac{f_{1}}{f_{1}\varepsilon_{\infty}+(1-f_{1})\varepsilon_{2}}}
\end{aligned},
\right.
\end{equation}
which stands for the location of one Dirac point in the $\omega$-$k$ space.
Since Eq.~\eqref{eq:tmm} is an even function of the wave vector $k_{y}$,
there are two Dirac points symmetrically distributed along $k_{y}$-direction
with respect to the origin.
It is interesting that Eq.~\eqref{eq:spp_sol} reveals that the positions of the
Dirac points in the band structure are only related on the filling ratios $f_{1}$
and $f_{2}$ of the materials in the multilayer stack, rather than the actual
thicknesses of different layers, which means that the positions of the Dirac points
are determined once the filling ratios of the materials are specified.
Moreover, the frequency $\omega/\omega_{p}=\sqrt{d_{1}/(\varepsilon_{\infty}d_{1}+\varepsilon_{2}d_{2})}$
is associated with the frequency at which the effective permittivity $\varepsilon_{y}^{(0)}=0$,
as calculated from Eq.~\eqref{eq:emt_y} based on the EMT.

It is necessary to investigate more about the dispersion relation around
the Dirac points with respect to the wave vector $k_{x}$ and $k_{y}$.
Figure~\blue{2(a)} shows the dispersion relation between the wave vector $k_{x}/k_{p}$
and the frequency $\omega/\omega_{p}$ calculated from Eq.~\eqref{eq:tmm} with
the wave vector
$k_{y}/k_{p}=\sqrt{\varepsilon_{2}d_{1}d_{2}/\left((d_{2}-d_{1})
(\varepsilon_{\infty}d_{1}+\varepsilon_{2}d_{2})\right)}$
around the Dirac point.
The band structure with the band~1 and the band~2 connected with the Dirac point is
clearly illustrated.
Furthermore, the linear dispersion at the Dirac point, a necessary condition
for a Dirac point, is clear revealed by the Taylor expansion, and indicated
by the black straight lines that obey the following equation
\begin{equation}
\label{eq:liner_x}
    \omega/\omega_{p} =
        \pm\frac{k_{p}(d_{1}+d_{2})}{\sqrt{2\Delta}}(k_{x}/k_{p})
        + \sqrt{\frac{d_{1}}{\varepsilon_{\infty}d_{1}+\varepsilon_{d}d_{2}}},
\end{equation}
with the constant $\Delta$, a function of parameters
$\varepsilon_{\infty}$, $\varepsilon_{2}$, $d_{1}$, and $d_{2}$, as listed in the Appendix.
Here the constant $\Delta$ has a value of $58.705$.
The linear dispersion relation between $k_{x}/k_{p}$ and $\omega/\omega_{p}$
forms the cross section of the Dirac cone at the Dirac point in the $\omega$-$k_{x}$ plane,
which is similar to the case in the electron band structure of graphene, implying that the
propagation of electromagnetic wave in the multilayer stack will be the optical analogue
to the charge transport in graphene.
In addition, Fig.~\blue{2(b)} displays the dispersion relation between the wave vector $k_{y}/k_{p}$
and the frequency $\omega/\omega_{p}$ based on Eq.~\eqref{eq:tmm} with the wave vector
$k_{x}/k_{p}=0$ around the Dirac point.
For comparison, the SPP dispersion relation
$k_{y}/k_{p}=\sqrt{\varepsilon_{1}\varepsilon_{2}/(\varepsilon_{1}+\varepsilon_{2})}\omega/\omega_{p}$,
and the dispersion relation based on the EMT are also plotted.
It is noted that the EMT dispersion relation is calculated from the following equation
\begin{equation}
\label{eq:ifc}
    \frac{k_{x}^{2}}{\varepsilon_{y}^{(0)}}
        + \frac{k_{y}^{2}}{\varepsilon_{x}^{(0)}}
        = \left(\frac{\omega}{c}\right)^{2},
\end{equation}
under the condition of the wave vector $k_{x}/k_{p}=0$, which implies two different bands as
\begin{equation}
\label{eq:emt_band_1}
    k_{y}/k_{p}=\sqrt{\varepsilon_{x}^{(0)}}\omega/\omega_{p},
\end{equation}
and
\begin{equation}
\label{eq:emt_band_2}
\left\{
\begin{aligned}
    & k_{y}/k_{p} \in \mathbb{R} \\
    & \omega/\omega_{p} = \sqrt{d_{1}/(\varepsilon_{\infty}d_{1}+\varepsilon_{2}d_{2})}
\end{aligned}\,,
\right.
\end{equation}
because $\varepsilon_{y}^{(0)}=0$ at the frequency of
$\omega/\omega_{p}=\sqrt{d_{1}/(\varepsilon_{\infty}d_{1}+\varepsilon_{2}d_{2})}$.
Clearly, the two bands of the dispersion curves from Eq.~\eqref{eq:tmm} for the
multilayer stack intersect at the Dirac point, and converge to the SPP dispersion
curve when the wave vector $k_{y}/k_{p}$ is increased, due to the surface plasmon
resonance (SPR) in the condition of $\varepsilon_{1}=-\varepsilon_{2}$.
On the contrary, the two bands of the dispersion curves based on the EMT of
Eqs.~\eqref{eq:emt_band_1} and \eqref{eq:emt_band_2} do not predict the SPR
at large wave vector $k_{y}/k_{p}$,
since the EMT does not take into account the layered structures of the multilayer
stack, but they still intersect at the Dirac point.
Furthermore, the dispersion relation from Eq.~\eqref{eq:tmm} for the multilayer
stack also obeys a linear relation at the Dirac point, which is indicated by the
black straight lines that obey the following equation
\begin{equation}
\label{eq:liner_y}
    \frac{\omega}{\omega_{p}}-\sqrt{\frac{d_{1}}{\varepsilon_{\infty}d_{1}+\varepsilon_{2}d_{2}}}
        = \frac{-C_{2}\pm\sqrt{C_{2}^{2}-4C_{1}C_{3}}}{2C_{1}}
        \left( \frac{k_{y}}{k_{p}}
        - \sqrt{\frac{\varepsilon_{2}d_{1}d_{2}}{(d_{2}-d_{1})
        (\varepsilon_{\infty}d_{1}+\varepsilon_{2}d_{2})}} \right).
\end{equation}
The constants $C_{1}$, $C_{2}$, and $C_{3}$ are the functions of parameters
$\varepsilon_{\infty}$, $\varepsilon_{2}$, $d_{1}$, and $d_{2}$, as listed in the Appendix.
Here the three constants have the values of $C_{1}=58.705$, $C_{2}=-20.0954$, and $C_{3}=0.282375$.
Equation~\eqref{eq:liner_y} reveals the asymmetry of the dispersion relation between
the wave vector $k_{y}/k_{p}$ and the frequency $\omega/\omega_{p}$
with respect to the Dirac point, which leads to a titled cross section
of the Dirac cone in the $\omega$-$k_{y}$ plane.
Additionally, it is noted that the location of the Dirac point is not at the center
of the BZ, where the wave vector $k_{x}/k_{p}=k_{y}/k_{p}=0$, due to the different
periods of the multilayer stack along $x$- and $y$-directions.
Therefore, the Dirac point cannot be mapped into an ENZ point in the $\omega$-$k$ space,
although the frequency of the Dirac point is the same as the ENZ frequency predicted
by the EMT of Eq.~\eqref{eq:emt_y}, where the effective permittivity $\varepsilon_{y}^{(0)}=0$.
Depicted in Fig.~\blue{2(c)}, the EMT predicted ENZ point is located at the position of
$k_{x}/k_{p}=k_{y}/k_{p}=0$
and $\omega/\omega_{p}=\sqrt{d_{1}/(\varepsilon_{\infty}d_{1}+\varepsilon_{2}d_{2})}$.
However, the nonlocal ENZ point for the multilayer stack due to the optical nonlocality
is shifted into the position of $k_{x}/k_{p}=k_{y}/k_{p}=0$ but with a lower frequency,
which is determined from Eq.~\eqref{eq:tmm}.

\section{Iso-frequency contours and propagating modes}

The IFCs of multilayer stack represents the spatial dispersion in $k$-space,
which are directly related to the propagating properties of electromagnetic waves.
Displayed in Fig.~\blue{3}, the IFCs of five different frequencies near the Dirac point
are plotted, including $636.577\,\mathrm{THz}$ [the nonlocal ENZ frequency from Eq.~\eqref{eq:tmm}],
$641\,\mathrm{THz}$ (slightly below the Dirac point),
$647.027\,\mathrm{THz}$ [at the Dirac point from Eq.~\eqref{eq:spp_sol}],
$651\,\mathrm{THz}$ (slightly above the Dirac point),
and $671\,\mathrm{THz}$ (far above the Dirac point).
The IFCs calculated from the band structure of Eq.~\eqref{eq:tmm} for the multilayer stack
are plotted as red curves, while the IFCs obtained from the EMT of Eq.~\eqref{eq:ifc}
are plotted as blue curves for comparison.
It is found that the IFCs obtained from Eq.~\eqref{eq:tmm} consist of two branches
at all frequencies, a hyperbolic-like branch and an elliptic-like branch, which is
coincident with the band structure shown in Fig.~\blue{1(c)}, and the two branches join
together at the Dirac point shown in Fig.~\blue{3(c)}.
On the contrary, the EMT calculated IFCs consist of only one single branch,
varying from hyperbolic-like to elliptic-like as the frequency is increased,
and they possess similar curvatures as that of the IFCs from Eq.~\eqref{eq:tmm}
when $\left|k_{x}/k_{p}\right|\ll1$, which is coincident with the long wavelength
limit for the EMT.
It is shown in Fig.~\blue{3(c)} that the EMT calculated IFC reduces into a straight
line along the $k_{y}$-axis at the frequency of the Dirac point, which is corresponding
to the band structure described by Eq.~\eqref{eq:emt_band_2}.
While the nonlocal IFC at the same frequency obtained from Eq.~\eqref{eq:tmm}
shows a dramatic difference, revealing a giant optical nonlocality at the Dirac point.
With the optical nonlocality, the permittivity will be a function of both the frequency
and the wave vector, which can be analytically described in an approximate way based on
a modification about the EMT of Eqs.~\eqref{eq:emt_x} and \eqref{eq:emt_y},
through the Taylor expansion of Eq.~\eqref{eq:tmm} under the conditions of
$\left|k_{x}(d_{1}+d_{2})\right|\ll1$,
$\left|k_{x}^{(1)}d_{1}\right|\ll1$,
and $\left|k_{x}^{(2)}d_{2}\right|\ll1$, which are all satisfied in the neighborhood
of the Dirac point.
By expanding Eq.~\eqref{eq:tmm} up to the third non-vanishing term,
the approximated dispersion relation can be written in the form of Eq.~\eqref{eq:ifc} as
\begin{equation}
\label{eq:nonlocal_ifc}
    \frac{k_{x}^{2}}{\varepsilon_{y}^{\mathrm{eff}}}
        +\frac{k_{y}^{2}}{\varepsilon_{x}^{\mathrm{eff}}}
        = \left(\frac{\omega}{c}\right)^{2},
\end{equation}
with the modified nonlocal effective permittivity components of
$\varepsilon_{x}^{\mathrm{eff}}=\varepsilon_{x}^{(0)}/(1-\delta_{x})$,
and $\varepsilon_{y}^{\mathrm{eff}}=\varepsilon_{y}^{(0)}/(1-\delta_{y})$,
where the nonlocal modification factors $\delta_{x}$ and $\delta_{y}$ are the functions
of both frequency and wave vector as
\begin{equation}
\label{eq:delta_x}
    \delta_{x} = \frac{A_{x1}+A_{x2}+A_{x3}+A_{x4}+A_{x5}+A_{x6}}
        {B_{x0}(B_{x1}+B_{x2}+B_{x3}+B_{x4}+B_{x5})},
\end{equation}
and
\begin{equation}
\label{eq:delta_y}
    \delta_{y} = \frac{A_{y1}+A_{y2}+A_{y3}+A_{y4}+A_{y5}}
        {B_{y1}+B_{y2}+B_{y3}+B_{y4}+B_{y5}},
\end{equation}
in which all the terms are listed in the Appendix.
The approximated IFC based on Eq.~\eqref{eq:nonlocal_ifc} at the frequency of
the Dirac point is plotted as black curves in Fig.~\blue{3(c)}, and it matches well
with the IFC directly obtained from Eq.~\eqref{eq:tmm} near the Dirac point.
The nonlocal modification factors clearly reveal that the modified nonlocal
effective permittivity $\varepsilon_{x}^{\mathrm{eff}}$ and $\varepsilon_{y}^{\mathrm{eff}}$
are not only related with the frequency, but also strongly dependent on the wave vector
components $k_{x}$ and $k_{y}$.
It is shown that at the Dirac point, the nonlocal modification factor $\delta_{x}=0$,
which means $\varepsilon_{x}^{\mathrm{eff}}=\varepsilon_{x}^{(0)}$ and the nonlocal
effect does not affect the effective permittivity component vertical to the multilayer
interface.
However, there is significant difference for the modified nonlocal effective permittivity
component $\varepsilon_{y}^{\mathrm{eff}}$, where the nonlocal modification factor
$\delta_{y}=1$ at the Dirac point, leading to an indeterminate form of the
$\varepsilon_{y}^{\mathrm{eff}}=\varepsilon_{y}^{(0)}/(1-\delta_{y})=0/0$,
since $\varepsilon_{y}^{(0)}=0$ at the Dirac point.
The limitation shows that
\begin{equation}
    \lim\left(\varepsilon_{y}^{\mathrm{eff}}\right) =
        \frac{\varepsilon_{2}^{2}d_{1}d_{2}^{2}k_{p}^{2}}
        {12(\varepsilon_{\infty}d_{1}+\varepsilon_{2}d_{2})} = 0.108129
\end{equation}
at the Dirac point, which is coincident with the previous analysis that the Dirac
point in the band structure of the multilayer stack cannot be mapped into the ENZ point.

According to the IFCs with nonlocal effects calculated from Eq.~\eqref{eq:tmm},
the corresponding eigenmodes and propagating patterns are analyzed for the
electromagnetic wave propagating along $y$-direction in the multilayer stack in
Fig.~\blue{4}, at three frequencies around the Dirac point including
$641\,\mathrm{THz}$, $647.027\,\mathrm{THz}$, and $651\,\mathrm{THz}$, displayed in Fig.~\blue{4}.
The eigenmodes supported in one period of
$\mathrm{Al}_2\mathrm{O}_3$-$\mathrm{Au}$-$\mathrm{Al}_2\mathrm{O}_3$
unit cell of the multilayer stack are illustrated with the distributions of the magnetic
field amplitude $H_{z}$ and the absolute value of the magnetic field $\mathrm{abs}(H_{z})$.
When the frequency is below or above the frequency of the Dirac point, the propagating
electromagnetic wave possesses two different eigenmodes, the symmetric mode and the
anti-symmetric mode, with different propagating constants $k_{y}$ as marked on the IFCs
in Fig.~\blue{3}.
The two eigenmodes degenerate into a single symmetric mode at the frequency of the Dirac point.
Besides, an interchange of the two eigenmodes occurs when the frequency pass across the frequency
of the Dirac point, due to a band inversion at the Dirac point as shown in Fig.~\blue{2(b)}.

In addition, the giant optical nonlocality near the Dirac point of the multilayer stack can
result in a unique optical phenomenon in the propagation of the electromagnetic wave.
The propagation of a TM polarized Gaussian beam (with $E_{x}$, $H_{z}$, and $k_{y}$)
is considered at the three different frequencies,
and the distributions of the absolute
value of the magnetic field $\mathrm{abs}(H_{z})$ are shown in Fig.~\blue{4}.
For comparison, both the multilayer stack and the corresponding effective medium are simulated.
Regarding the normal incidence, it is found that the Gaussian beam is scattered into the similar
diverging patterns in both multilayer stack and effective medium, when the frequency is below
and above the frequency of the Dirac point, due to the common sharp curvature of the IFCs near
the Dirac point in Figs.~\blue{3(b)} and \blue{3(d)}.
However, at the frequency of the Dirac point, the propagation of the electromagnetic wave
is extraordinary.
In the multilayer stack, due to the degeneration of the symmetric mode and the anti-symmetric
mode, the joint of the two IFC branches at the Dirac point flatten the sharp curvature, leading
to a splitting of the Gaussian beam into two mirrored propagating directions, as shown in Fig.~\blue{4(b)}.
The beam splitting phenomenon in the multilayer stack due to the optical nonlocality can be
looked as the optical analogue to the giant nonlocality enhanced long-range flavor current
splitting in the quantum Hall effect of graphene at the Dirac point.
On the contrary, the incident electromagnetic wave is totally prevented from propagating into
the corresponding effective medium due to the large impedance mismatch, which is coincident
with the EMT calculated IFC shown in Fig.~\blue{3(c)}.

\section{Conclusions}

The giant optical nonlocality near the Dirac point for the multilayer stack have
been revealed in lossless metal-dielectric multilayer metamaterials by studying
the band structure in the three dimensional $\omega$-$k$ space based on the
transfer-matrix method with the optical nonlocal effect.
The exact location of the Dirac point in the band structure of the multilayer
stack is determined and the dispersion relation around the Dirac point is investigated
in details.
Based on the mode analysis, it is proved that the degeneracy of the symmetric mode
and the asymmetric mode of the coupled SPPs is the origin of the Dirac point.
Meanwhile, the position shift of the ENZ point in the band structure affected by
the giant optical nonlocality near the Dirac point is also explored, and a nonlocal
modification on the dispersion relation based on the EMT including the optical
nonlocality near the Dirac point is derived.
Furthermore, the giant optical nonlocality near the Dirac point is revealed by means
of the IFC analysis, and the extraordinary beam splitting at the Dirac point induced
by the giant optical nonlocality is also demonstrated.
Finally, it is noted that although the study is carried out under lossless condition,
the giant optical nonlocality still affects the propagation of the electromagnetic wave
with a moderate material loss.

\section*{Appendix}

The constant $\Delta$ in the linear dispersion relation Eq.~\eqref{eq:liner_x} between
the wave vector $k_{x}/k_{p}$ and the frequency $\omega/\omega_{p}$ can be calculated as follows
\begin{equation}
    \Delta = -\frac{1}{4\varepsilon_{2}^{2}d_{1}^{3}d_{2}^{4}}
        \left(\Delta_{1}+\Delta_{2}+\Delta_{3}+\Delta_{4}\right),
\end{equation}
in which
\begin{equation}
\left\{
\begin{aligned}
    & \Delta_{1} = \varepsilon_{\infty}\varepsilon_{2}d_{1}d_{2}
        (\varepsilon_{\infty}d_{1}^{2}+\varepsilon_{2}d_{2}^{2})
        (d_{1}+d_{2})^{3} \\
    & \Delta_{2} = (\varepsilon_{\infty}d_{1}^{2}+\varepsilon_{2}d_{2}^{2})
        (\varepsilon_{\infty}^{2}d_{1}^{3}+\varepsilon_{2}^{2}d_{2}^{3})
        (d_{1}+d_{2})^{2} \\
    & \Delta_{3} = 2\varepsilon_{2}d_{1}^{2}d_{2}^{2}k_{p}^{2}
        (\varepsilon_{\infty}d_{1}^{2}-\varepsilon_{2}d_{2}^{2})
        (d_{2}-d_{2}) \\
    & \Delta_{4} = -(\varepsilon_{\infty}d_{1}+\varepsilon_{2}d_{2})
        (\varepsilon_{\infty}d_{1}^{2}+\varepsilon_{2}d_{2}^{2})^{2}
        (d_{1}+d_{2})^{2}
        \cosh\left(\frac{2\sqrt{\varepsilon_{2}}d_{1}d_{2}k_{p}}
        {\sqrt{(d_{2}-d_{1})(\varepsilon_{\infty}d_{1}+\varepsilon_{2}d_{2})}}\right)
\end{aligned}.
\right.
\end{equation}

The constants $C_{1}$, $C_{2}$, and $C_{3}$ in the linear dispersion relation
Eq.~\eqref{eq:liner_y} between the wave vector $k_{y}/k_{p}$ and the frequency
$\omega/\omega_{p}$ can be calculated as follows
\begin{equation}
    C_{1} = -\frac{1}{4\varepsilon_{2}^{2}d_{1}^{3}d_{2}^{4}}
        \left( C_{1}^{(1)}+C_{1}^{(2)}+C_{1}^{(3)}+C_{1}^{(4)}+C_{1}^{(5)} \right),
\end{equation}
\begin{equation}
    C_{2} = -\frac{d_{1}^{2}-d_{2}^{2}}{2\varepsilon_{2}d_{1}^{3}d_{2}^{3}}
        \sqrt{\frac{d_{2}-d_{1}}{\varepsilon_{2}d_{2}}}
        \left( C_{2}^{(1)}+C_{2}^{(2)}+C_{2}^{(3)}+C_{2}^{(4)} \right),
\end{equation}
and
\begin{equation}
    C_{3} = -\frac{(d_{1}^{2}-d_{2}^{2})^{2}}{4\varepsilon_{2}d_{1}^{3}d_{2}^{3}}
        \left( C_{3}^{(1)}+C_{3}^{(2)}+C_{3}^{(3)} \right),
\end{equation}
with all terms listed as follows
\begin{equation}
\left\{
\begin{aligned}
    & C_{1}^{(1)} = \varepsilon_{2}^{3}d_{2}^{5}
        \left( (d_{1}+d_{2})^{2} - 2d_{1}^{2}d_{2}(d_{1}-d_{2})k_{p}^{2} \right) \\
    & C_{1}^{(2)} = \varepsilon_{\infty}\varepsilon_{2}^{2}d_{1}d_{2}^{3}
        \left( (d_{1}+d_{2})^{2}(2d_{1}+d_{2}) + 4d_{1}^{3}d_{2}(d_{1}-d_{2})k_{p}^{2} \right) \\
    & C_{1}^{(3)} = \varepsilon_{\infty}^{2}\varepsilon_{2}d_{1}^{3}d_{2}
        \left( (d_{1}+d_{2})^{2}(d_{1}+2d_{2}) - 2d_{1}^{3}d_{2}(d_{1}-d_{2})k_{p}^{2} \right) \\
    & C_{1}^{(4)} = \varepsilon_{\infty}^{3}d_{1}^{5}(d_{1}+d_{2})^{2} \\
    & C_{1}^{(5)} = -(d_{1}+d_{2})^{2}
        (\varepsilon_{\infty}d_{1}+\varepsilon_{2}d_{2})
        (\varepsilon_{\infty}d_{1}^{2}+\varepsilon_{2}d_{2}^{2})^{2}
        \cosh\left(\frac{2\sqrt{\varepsilon_{2}}d_{1}d_{2}k_{p}}
        {\sqrt{(d_{2}-d_{1})(\varepsilon_{\infty}d_{1}+\varepsilon_{2}d_{2})}}\right)
\end{aligned},
\right.
\end{equation}
\begin{equation}
\left\{
\begin{aligned}
    & C_{2}^{(1)} = \varepsilon_{2}^{2}d_{2}^{3}
        \left( d_{1}+d_{2}+2d_{1}^{2}d_{2}k_{p}^{2} \right) \\
    & C_{2}^{(2)} = \varepsilon_{\infty}\varepsilon_{2}d_{1}d_{2}
        \left( (d_{1}+d_{2})^{2} - 2d_{1}^{3}d_{2}k_{p}^{2} \right) \\
    & C_{2}^{(3)} = \varepsilon_{\infty}^{2}d_{1}^{3}
        \left( d_{1}+d_{2} \right) \\
    & C_{2}^{(4)} = -(d_{1}+d_{2})
        (\varepsilon_{\infty}d_{1}+\varepsilon_{2}d_{2})
        (\varepsilon_{\infty}d_{1}^{2}+\varepsilon_{2}d_{2}^{2})
        \cosh\left(\frac{2\sqrt{\varepsilon_{2}}d_{1}d_{2}k_{p}}
        {\sqrt{(d_{2}-d_{1})(\varepsilon_{\infty}d_{1}+\varepsilon_{2}d_{2})}}\right)
\end{aligned},
\right.
\end{equation}
and
\begin{equation}
\left\{
\begin{aligned}
    & C_{3}^{(1)} = \varepsilon_{2}d_{2}
        \left( -d_{1}+d_{2}+2d_{1}^{2}d_{2}k_{p}^{2} \right) \\
    & C_{3}^{(2)} = \varepsilon_{\infty}d_{1}
        \left( -d_{1}+d_{2} \right) \\
    & C_{3}^{(3)} = (d_{1}-d_{2})
        (\varepsilon_{\infty}d_{1}+\varepsilon_{2}d_{2})
        \cosh\left(\frac{2\sqrt{\varepsilon_{2}}d_{1}d_{2}k_{p}}
        {\sqrt{(d_{2}-d_{1})(\varepsilon_{\infty}d_{1}+\varepsilon_{2}d_{2})}}\right)
\end{aligned}.
\right.
\end{equation}

The modification factors $\delta_{x}$ and $\delta_{y}$ in
Eqs.~\eqref{eq:delta_x} and \eqref{eq:delta_y} are based on the following terms
\begin{equation}
\left\{
\begin{aligned}
    & A_{x1} = \varepsilon_{1}\varepsilon_{2}
        \left( k_{y}^{2} - \varepsilon_{1}k^{2} \right)d_{1}^{5} \\
    & A_{x2} = \left(
        (2\varepsilon_{1}^{2}+\varepsilon_{1}\varepsilon_{2}+2\varepsilon_{2}^{2})k_{y}^{2}
        - \varepsilon_{1}(\varepsilon_{1}^{2}+2\varepsilon_{1}\varepsilon_{2}+2\varepsilon_{2}^{2})k^{2}
        \right)d_{1}^{4}d_{2} \\
    & A_{x3} = 2\left(
        (\varepsilon_{1}^{2}+3\varepsilon_{1}\varepsilon_{2}+\varepsilon_{2}^{2})k_{y}^{2}
        - \varepsilon_{1}\varepsilon_{2}(3\varepsilon_{1}+2\varepsilon_{2})k^{2}
        \right)d_{1}^{3}d_{2}^{2} \\
    & A_{x4} = 2\left(
        (\varepsilon_{1}^{2}+3\varepsilon_{1}\varepsilon_{2}+\varepsilon_{2}^{2})k_{y}^{2}
        - \varepsilon_{1}\varepsilon_{2}(2\varepsilon_{1}+3\varepsilon_{2})k^{2}
        \right)d_{1}^{2}d_{2}^{3} \\
    & A_{x5} = \left(
        (2\varepsilon_{1}^{2}+\varepsilon_{1}\varepsilon_{2}+2\varepsilon_{2}^{2})k_{y}^{2}
        - \varepsilon_{1}(2\varepsilon_{1}^{2}+2\varepsilon_{1}\varepsilon_{2}+\varepsilon_{2}^{2})k^{2}
        \right)d_{1}d_{2}^{4} \\
    & A_{x6} = \varepsilon_{1}\varepsilon_{2}\left( k_{y}^{2}-\varepsilon_{2}k^{2} \right)d_{2}^{5}
\end{aligned},
\right.
\end{equation}
\begin{equation}
\left\{
\begin{aligned}
    & B_{x0} = \varepsilon_{1}d_{2}+\varepsilon_{2}d_{1} \\
    & B_{x1} = \varepsilon_{1}^{2}k^{2}d_{1}^{4} \\
    & B_{x2} = 2\varepsilon_{1}(\varepsilon_{1}+\varepsilon_{2})k^{2}d_{1}^{3}d_{2} \\
    & B_{x3} = 6\varepsilon_{1}(\varepsilon_{2}k^{2}d_{2}^{2}-2)d_{1}^{2} \\
    & B_{x4} = 2(\varepsilon_{1}+\varepsilon_{2})
        (\varepsilon_{2}k^{2}d_{2}^{2}-6)d_{1}d_{2} \\
    & B_{x5} = \varepsilon_{2}(\varepsilon_{2}k^{2}d_{2}^{2}-12)d_{2}^{2}
\end{aligned},
\right.
\end{equation}
and
\begin{equation}
\left\{
\begin{aligned}
    & A_{y1} = \varepsilon_{1}\left( \varepsilon_{1}k^{2} - k_{x}^{2} \right)d_{1}^{4} \\
    & A_{y2} = \left( 2\varepsilon_{1}(\varepsilon_{1}+\varepsilon_{2})k^{2}
        - (3\varepsilon_{1}+\varepsilon_{2})k_{x}^{2} \right)d_{1}^{3}d_{2} \\
    & A_{y3} = 3\left( 2\varepsilon_{1}\varepsilon_{2}k^{2}
        - (\varepsilon_{1}+\varepsilon_{2})k_{x}^{2} \right)d_{1}^{2}d_{2}^{2} \\
    & A_{y4} = \left( 2\varepsilon_{2}(\varepsilon_{1}+\varepsilon_{2})k^{2}
        - (\varepsilon_{1}+3\varepsilon_{2})k_{x}^{2} \right)d_{1}d_{2}^{3} \\
    & A_{y5} = \varepsilon_{2}\left( \varepsilon_{2}k^{2} - k_{x}^{2} \right)d_{2}^{4}
\end{aligned},
\right.
\end{equation}
\begin{equation}
\left\{
\begin{aligned}
    & B_{y1} = \varepsilon_{1}^{2}k^{2}d_{1}^{4} \\
    & B_{y2} = 2\varepsilon_{1}(\varepsilon_{1}+\varepsilon_{2})k^{2}d_{1}^{3}d_{2} \\
    & B_{y3} = 6\varepsilon_{1}\left( \varepsilon_{2}k^{2}d_{2}^{2}-2 \right)d_{1}^{2} \\
    & B_{y4} = 2(\varepsilon_{1}+\varepsilon_{2})
        \left( \varepsilon_{2}k^{2}d_{2}^{2}-6 \right)d_{1}d_{2} \\
    & B_{y5} = \varepsilon_{2}
        \left( \varepsilon_{2}k^{2}d_{2}^{2}-12 \right)d_{2}^{2}
\end{aligned},
\right.
\end{equation}

\section*{Acknowledgments}

This work was partially supported by the Department of Mechanical and Aerospace Engineering,
the Materials Research Center, the Intelligent Systems Center, and the Energy Research and
Development Center at Missouri S\&T, the University of Missouri Research Board,
and the Ralph E. Powe Junior Faculty Enhancement Award.
The authors acknowledge S. Feng for some useful discussions about this work.



\newpage                  %
\section*{Figure Captions}%
\noindent
\textbf{FIG.~1}.
    (a) The metal-dielectric multilayer stack consists of alternating layers of gold ($\mathrm{Au}$)
    and alumina ($\mathrm{Al}_2\mathrm{O}_3$), with $d_{1}=15\,\mathrm{nm}$ and $d_{2}=35\,\mathrm{nm}$ and
    the permittivity $\varepsilon_{1}$ and $\varepsilon_{2}$, respectively.
    (b) The band structure of the multilayer stack in the first BZ calculated from Eq.~\eqref{eq:tmm}
    in three dimensional $\omega$-$k$ space.
    (c) The band~1 and band~2 in the band structure of the multilayer stack is connected
    by two Dirac points at the positions determined by Eq.~\eqref{eq:spp_sol}.

\vspace{5.0mm}
\noindent
\textbf{FIG.~2}.
    (a) The dispersion relation $k_{x}/k_{p}\sim\omega/\omega_{p}$  based on Eq.~\eqref{eq:tmm},
    under the condition of $k_{y}/k_{p}=\sqrt{\varepsilon_{2}d_{1}d_{2}/\left( (d_{2}-d_{1})
    (\varepsilon_{\infty}d_{1}+\varepsilon_{2}d_{2}) \right)}$ in red curves.
    The band~1 and band~2 are connected at the Dirac point,
    located at the position of $k_{x}/k_{p}=0$ according to Eq.~\eqref{eq:spp_sol}.
    The black lines indicate the linear dispersion in the neighborhood of
    the Dirac point consistent with Eq.~\eqref{eq:liner_x}.
    (b) The dispersion relation $k_{y}/k_{p}\sim\omega/\omega_{p}$
    near the Dirac point based on Eq.~\eqref{eq:tmm},
    with respect to $k_{x}/k_{p}=0$ in red curves.
    The dispersion relation obtained from the EMT and the SPP dispersion relation
    are plotted in dot-dashed black curve and dashed blue curves, respectively.
    All the dispersion curves intersect at the Dirac point.
    The dispersion curves obtained from the multilayer stack converge to the
    SPP dispersion when the wave vector $k_{y}/k_{p}$ increases, due to the SPR.
    The linear dispersion relation in the neighborhood of the Dirac point is
    plotted in black lines based on Eq.~\eqref{eq:liner_y}.
    (c) The positions of the ENZ determined from the EMT and the multilayer
    stack including the optical nonlocality are marked for comparison.

\vspace{5.0mm}
\noindent
\textbf{FIG.~3}.
    The variations of the IFCs at five different frequencies around the Dirac point
    (a) $636.577\,\mathrm{THz}$ (the nonlocal ENZ frequency),
    (b) $641\,\mathrm{THz}$ (slightly below the Dirac point),
    (c) $647.027\,\mathrm{THz}$ (at the Dirac point),
    (d) $651\,\mathrm{THz}$ (slightly above the Dirac point), and
    (e) $671\,\mathrm{THz}$ (far above the Dirac point).
    The IFCs from Eq.~\eqref{eq:tmm} with the optical nonlocality are plotted in red curves,
    while the EMT calculated IFCs are plotted in blue curves.
    The IFC of air is plotted in green circle for reference.
    The IFCs from Eq.~\eqref{eq:tmm} consist of two branches, and two eigenmodes degenerate
    at the Dirac point.
    On the contrary, only a single branch exists in the EMT calculated IFCs,
    which reduces into a straight line at the frequency of the Dirac point.

\vspace{5.0mm}
\noindent
\textbf{FIG.~4}.
    The eigenmodes and the propagating patterns of the electromagnetic wave in
    the multilayer stack at three different frequencies of
    (a) $641\,\mathrm{THz}$,
    (b) $647.027\,\mathrm{THz}$, and
    (c) $651\,\mathrm{THz}$
    near the Dirac point, corresponding to the IFCs in Fig.~\blue{3}.
    The symmetry eigenmode and the asymmetry eigenmode in one
    $\mathrm{Al}_{2}\mathrm{O}_{3}$-$\mathrm{Au}$-$\mathrm{Al}_{2}\mathrm{O}_{3}$
    unit cell of the multilayer stack are represented by the amplitude and the absolute
    value of the magnetic field $H_{z}$.
    The two eigenmodes degenerate as one symmetric mode at the frequency of the Dirac point,
    and invert as the frequency across the frequency of the Dirac point.
    The propagating patterns are plotted for both the multilayer stack and the corresponding
    effective medium for a normal incident Gaussian beam of TM mode ($E_{x}$, $H_{z}$, and $k_{y}$),
    represented by the distribution of the absolute value of the magnetic field $\mathrm{abs}(H_{z})$.
    Similar diverging patterns can be observed when the frequency is above and below the Dirac point,
    due to the sharp curvature of the IFC.
    The giant optical nonlocality at the Dirac point leads to a beam splitting
    in the multilayer stack that is dramatically different from the propagating
    property in the effective medium.


\clearpage
\newpage
\begin{figure}[htbp]
\renewcommand\thefigure{1}
    \centering
    \includegraphics[width=8.5cm]{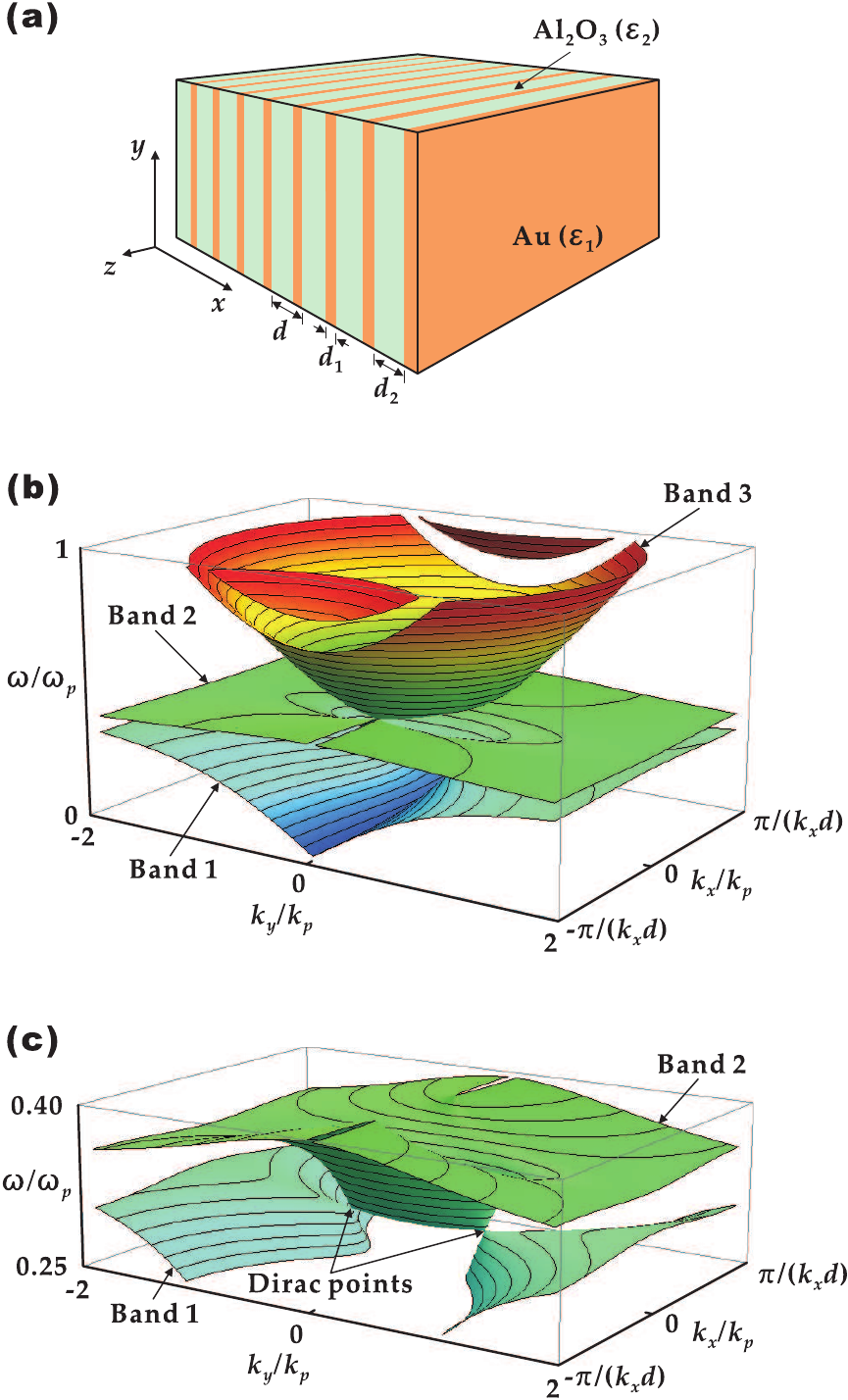}
    \caption{}
    \label{fig:fig1}
\end{figure}

\newpage
\begin{figure}[htbp]
\renewcommand\thefigure{2}
    \centering
    \includegraphics[width=7.0cm]{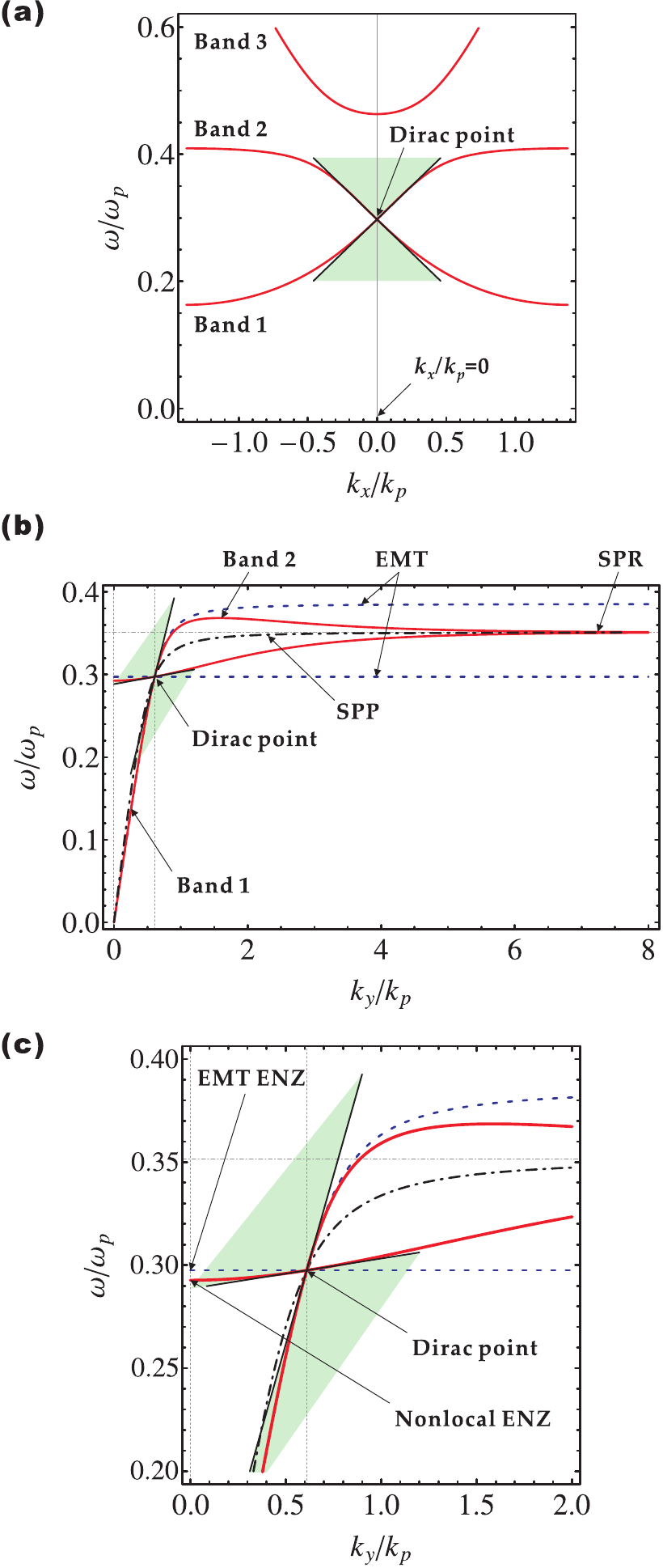}
    \caption{}
    \label{fig:fig2}
\end{figure}

\newpage
\begin{figure}[htbp]
\renewcommand\thefigure{3}
    \centering
    \includegraphics[width=12.5cm]{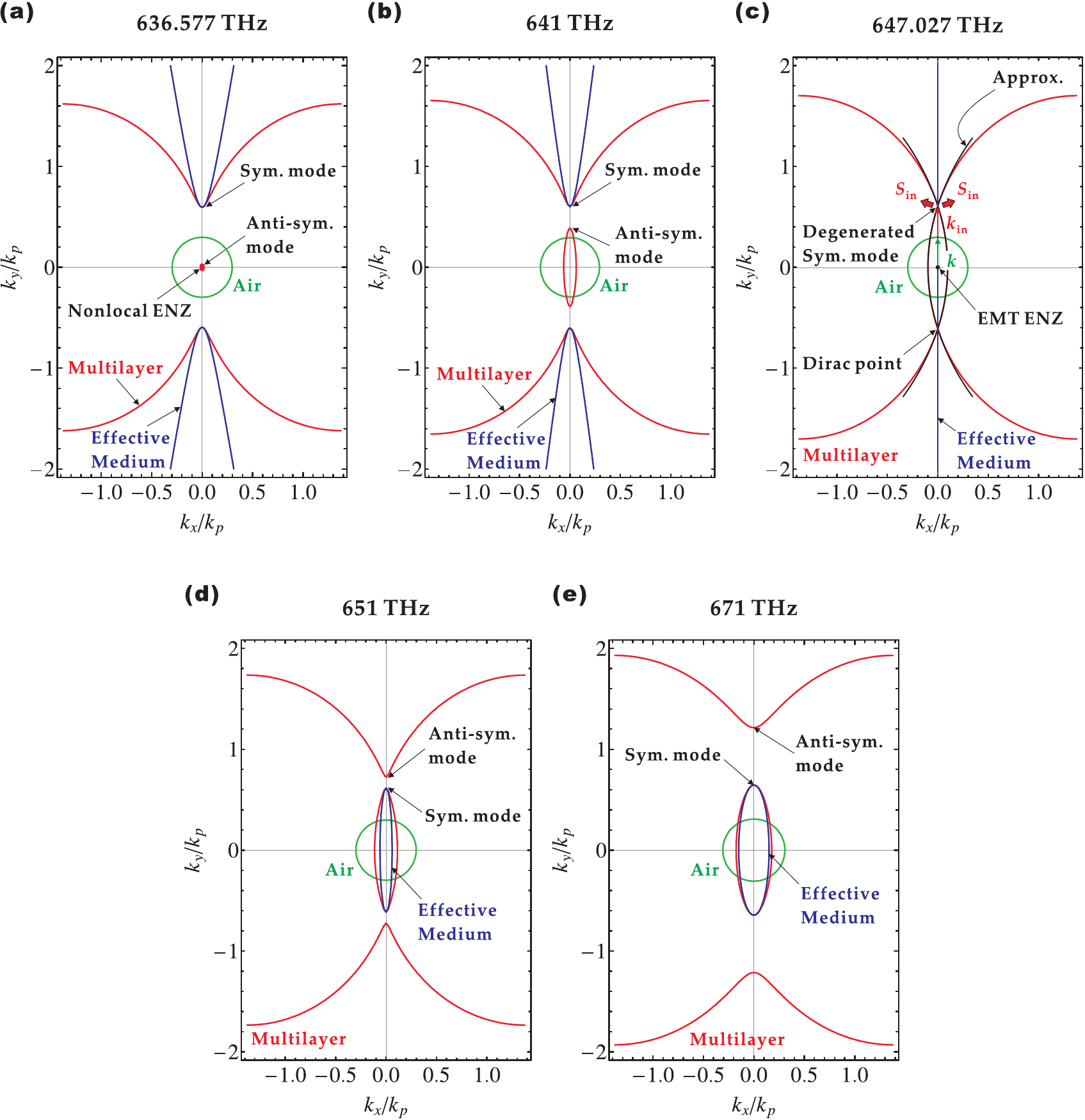}
    \caption{}
    \label{fig:fig3}
\end{figure}

\newpage
\begin{figure}[htbp]
\renewcommand\thefigure{4}
    \centering
    \includegraphics[width=12.5cm]{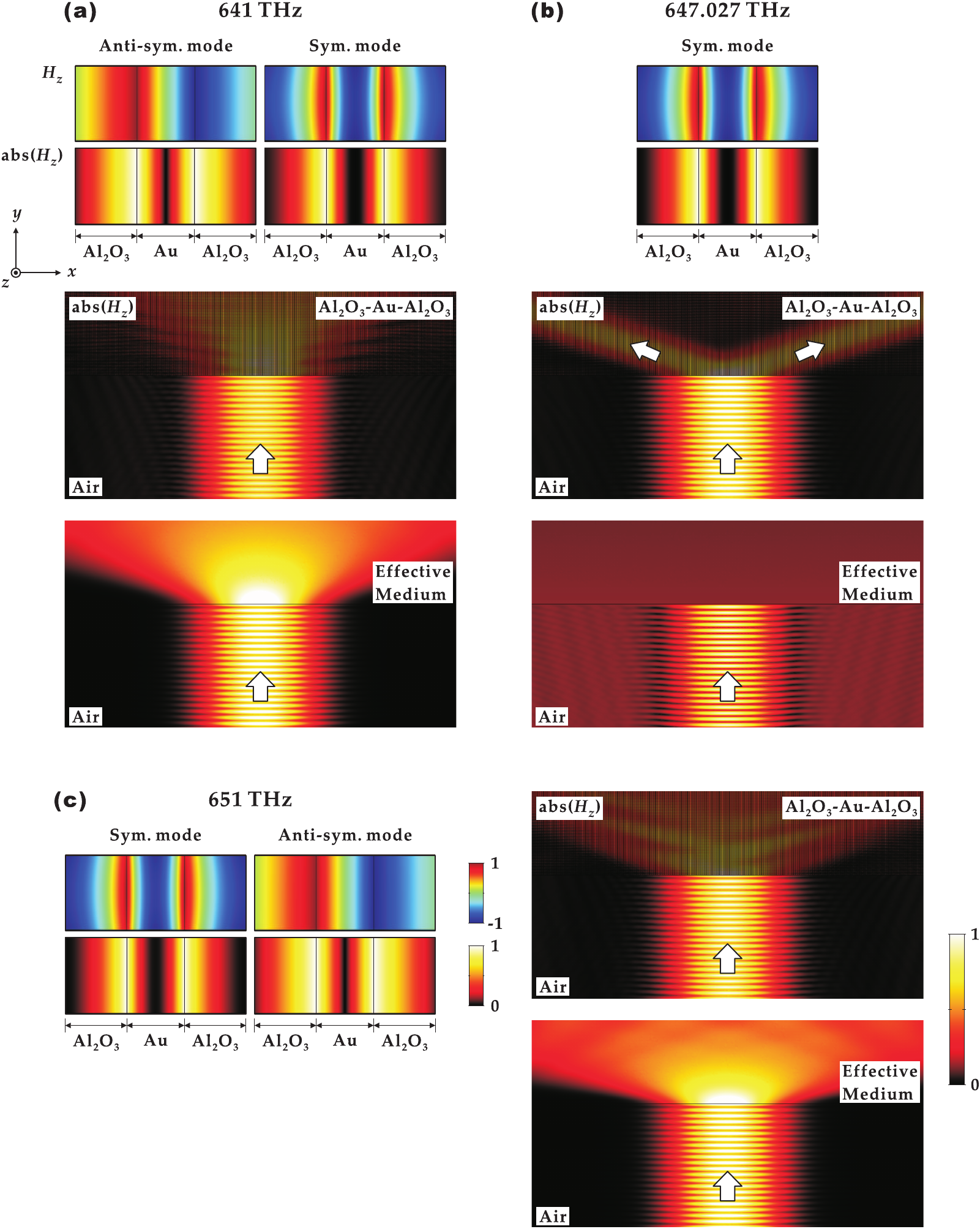}
    \caption{}
    \label{fig:fig4}
\end{figure}

\end{document}